\documentclass[letterpaper]{article}
\usepackage[preprint]{aaai2027}
\usepackage[hyphens]{url}
\usepackage{graphicx}
\usepackage{natbib}
\usepackage{caption}
\usepackage{algorithm}
\usepackage{algorithmic}
\usepackage{booktabs}
\usepackage{amsmath}
\usepackage{amssymb}
\usepackage{amsfonts}
\usepackage{amsthm}
\newtheorem{proposition}{Proposition}

\title{Characterizing and Defending Utility-Hijacking in Instruction-Tuning Data Selection}
\author{
Yahao Ren\textsuperscript{1},
Jian Lou\textsuperscript{1},
Huadi Zheng\textsuperscript{2},
Jiawei Duan\textsuperscript{1,*}
}
\affiliations{
\textsuperscript{1}Sun Yat-sen University, Guangzhou, China\\
\textsuperscript{2}Huawei Technologies Co., Ltd, Shenzhen, China\\
2023005786@link.tyut.edu.cn, jian.lou@hoiying.net\\
zhenghuadi@huawei.com, duanjw6@mail.sysu.edu.cn\\
\textsuperscript{*}Corresponding author.
}

\begin{document}
\maketitle
\begin{abstract}
Instruction following, a language model's ability to carry out user instructions, is a core capability whose quality depends heavily on the instruction-tuning data. To curate this data, an \emph{instruction-data selector} chooses which instruction-following records are used for fine-tuning, improving the model's alignment quality, raising training efficiency, and filtering out harmful records. A selector's choices thus directly affect the fine-tuned model's safety, and prior work shows that fine-tuning on unsafe data can even break the safety of an already safety-aligned model. However, despite their widespread use, the security of these selectors remains under-studied.

We conduct the first systematic study of adaptive supplier-side attacks against utility-oriented instruction-data selectors and introduce \emph{utility-hijacking}. Utility-hijacking shapes harmful records to remain compatible with a selector while preserving their malicious payload, causing top-$k$ selection to concentrate rather than filter the poison. Achieving both is non-trivial, because optimizing a record for score tends to wash out its payload, while bluntly preserving the payload makes the record look low-value. Our supplier-side generate-and-rerank procedure generates, per instruction, many payload-bearing candidates (for payload preservation) and submits only the highest-scoring survivor (for selector compatibility). Evaluated across three payloads (over-refusal, false-fact injection, and insecure code), six selectors, and victim models from Qwen2.5-7B to Qwen3-32B, utility-hijacking concentrates poison on vulnerable selector-payload pairs and changes model behavior, driving refusal from 0\% to 94\% and insecure code up to 56/90, while false-fact enrichment intensifies with scoring-model scale from 2.84$\times$ to 7.04$\times$. To defend against this attack, we propose \emph{CONCERT}, a targeted defense needing no content detector or poison labels. It flags suppliers selected far more often than chance allows, then re-selects under the same budget, keeping only records that two dissimilar selectors both rank highly. Across two payloads, four selectors, and four evaluation benchmarks on a Qwen3-32B victim, CONCERT reverses the poison concentration from 6.67$\times$ to 0$\times$ (code) and 3.59$\times$ to 0.67$\times$ (facts), and reduces downstream harm by 100\% and 75\% respectively, at unchanged capability.
\end{abstract}

\section{Introduction}

Building a strong instruction-following model increasingly means training on data the owner did not create. Owners gather instruction--response records from public datasets, synthetic generators, and data vendors \citep{ouyang2022training,wang2023selfinstruct,taori2023stanford,xu2023wizardlm}, and because training on all of them is costly and can even hurt quality, they pass the pool to an automated \emph{instruction-data selector} that scores every record and fine-tunes only on a fixed-size, top-scoring subset \citep{xia2024lessselecting,liu2024makesgood,li2024fromquantity,chen2024alpagasus}. A selector is valued because it raises the alignment quality of the tuned model, cuts training cost, and is trusted to leave low-value or harmful records behind. Yet its choices decide what the model learns, so the safety of the admitted data carries into the fine-tuned model, and prior work on harmful fine-tuning shows that unsafe records can even undo the safety of an already aligned model \citep{qi2024finetuning}. Despite how widely selectors are now deployed, their own security has barely been examined.

This very component, however, can be turned against the model owner. If an adversarial supplier can make harmful records, which we hereafter call \emph{poison}, look valuable to the selector, then top-$k$ selection will not filter them out. It will \emph{concentrate} them, spending a larger share of the fixed training budget on exactly the data the attacker wants. We call this failure mode \emph{utility-hijacking}, since the attacker hijacks the selector's notion of training utility rather than breaking the selector, the training code, or the evaluation.

Utility-hijacking sits in a blind spot between two lines of work. Poisoning studies typically assume the poison already reaches training \citep{wan2023poisoninglanguage,shu2023exploitabilityinstruction,qi2024finetuning}, whereas data-selection studies typically assume benign suppliers, so neither asks what happens when a supplier \emph{adapts} records to survive a utility selector. A successful attack has to reconcile two objectives that pull against each other, \emph{selector compatibility} so the poison scores highly enough to be selected, and \emph{payload preservation} so it still carries harm once training is done. Pushing a record's score tends to wash out its payload, which is what separates utility-hijacking from appending a trigger to otherwise clean data. Our supplier-side construction, \emph{generate-and-rerank}, fixes the payload and searches only for compatibility.

What has been missing is any account of \emph{when} utility-hijacking works. Top-$k$ selection keeps only the \emph{upper tail} of the score distribution, so if a payload reaches that tail at a higher rate than clean data, the very act of keeping only the best turns filtering into amplification. We formalize this as \textbf{CREST} (Compatibility-driven Rank Enrichment in Selector Tails), an exact closed form for poison concentration together with a statistic that separates vulnerable from immune pairs with no overlap in our experiments (every vulnerable pair $\ge0.596$, every immune pair $\le0.008$).

We study three unrelated, alignment-relevant payloads, namely over-refusal (helpfulness), false facts (honesty), and insecure code (safety). Across six selectors and victim models from Qwen2.5-7B to Qwen3-32B, utility-hijacking is both selector- and payload-dependent, yet on vulnerable pairs it is severe. Within the paired IFD cell, adaptation lifts poison concentration from $1.21\times$ to $6.07\times$, drives refusal from $0\%$ to $94\%$, elicits insecure code in up to $56/90$ cases, and intensifies with scoring-model scale from $2.84\times$ to $7.04\times$. It also does not require the attacker to supply much of the pool, since poison still concentrates $10.78\times$ at a $1\%$ input rate. Immunity does not transfer across payloads, which rules out simply ``using a safer selector'' and points at the structure a defense needs.

\textbf{CONCERT} (Consensus-Certified Selection) is our targeted answer, a defense that trusts no single selector and needs neither a content detector nor poison labels. A \emph{provenance audit} flags suppliers over-selected beyond chance, and a \emph{consensus reselection} keeps the same fixed budget but admits only records ranked highly by \emph{two} decorrelated selectors. Because poison shaped for one selector's tail is usually mediocre to the other, this rule carries a deterministic, label-free rank-intersection certificate. On a Qwen3-32B victim it drives concentration from $3.59\times$ to $0.67\times$ for false facts and $6.67\times$ to $0\times$ for insecure code, removes unsafe-code outputs ($56/90$ to $0/90$), and keeps most of the high-value clean data an immune selector would discard.

In summary, this paper makes the following contributions.
\begin{itemize}
  \item \textbf{A new attack on data selectors (utility-hijacking).} To our knowledge, this is the first systematic study of adaptive supplier-side attacks against utility-oriented instruction-data selectors. A supplier-side generate-and-rerank construction shapes poison to satisfy selector compatibility and payload preservation at once, so top-$k$ selection concentrates it rather than filtering it out.
  \item \textbf{An upper-tail account of utility-hijacking (CREST).} CREST reduces selection-stage poisoning to upper-tail survival, gives an exact closed form for poison concentration in terms of two survival probabilities, and provides a statistic that separates vulnerable from immune selector--payload pairs after scoring but before any model is trained.
  \item \textbf{A rank-intersection defense (CONCERT).} We combine an exact-test provenance audit with error-rate control and a fixed-budget consensus rule whose set-containment certificate excludes any record outside either selector's certified prefix, without any poison labels. The defense still holds when a white-box attacker reranks against the consensus objective itself, and CREST predicts why.
\end{itemize}

\section{Related Work}

\paragraph{Data selection for instruction tuning.}
Instruction tuning keeps only a small high-value subset of a large pool, scoring records by influence \citep{xia2024lessselecting,dai2025improvinginfluence}, difficulty \citep{li2024fromquantity,li2024superfiltering}, quality \citep{liu2024makesgood,chen2024alpagasus,li2024nuggets}, or tag diversity \citep{lu2024instag}, with small curated sets matching training on everything \citep{zhou2023lima,ivison2025largescale,zhang2025surveydata}. This line evaluates selectors under benign suppliers. We show the same selectors are an integrity boundary an adaptive supplier can invert, so a stronger utility selector is not automatically safer.

\paragraph{Poisoning, backdoors, and adaptive poison.}
Poisoning and backdoors implant harmful behavior through the training set, inserting triggers or corrupted records directly \citep{gu2017badnets,kurita2020weight,wallace2021concealed} or steering a model with a few instruction-tuning examples \citep{wan2023poisoninglanguage,shu2023exploitabilityinstruction,xu2024instructionsas,yan2024backdooringinstruction}, even at near-constant scale \citep{souly2025nearconstant,carlini2024poisoning,fu2024poisonbench}, and others optimize the poison for potency or stealth against a detector \citep{qiang2024learningpoison,zhang2024advbdgen,hubinger2024sleeper}. All assume the poison reaches training. We add a prior constraint, that it must first survive a utility selector, which makes selector compatibility rather than trigger design the objective.

\paragraph{Filtering, auditing, and robust selection.}
Defenses either harden the selector for safety \citep{shen2025sealsafety,li2025layeraware,he2024safedata,bianchi2024safety} or filter individual records, which has fundamental limits once the adversary adapts \citep{davies2025fundamentallimitations,huang2024harmfulsurvey}. Provenance is normally used for licensing \citep{longpre2024provenance} and data value for pricing \citep{ghorbani2019datashapley}. CONCERT instead turns provenance into an exact over-selection test with multiple-testing control (Holm FWER and BH FDR) \citep{holm1979simple,benjamini1995controlling} and combines two decorrelated selectors by rank consensus \citep{dwork2001rank}, giving a label-free set-containment certificate rather than a pointwise filter.

\section{Problem Formalization and Threat Model}

We study a \emph{select-then-train} pipeline in which a model owner (the \emph{victim}) holds a clean pool and also accepts records from external \emph{suppliers}, and ask what happens when one supplier is adversarial.

\paragraph{The pipeline.}
Let the clean pool be $\mathcal{D}_{\mathrm{clean}}$, and let an attacker contribute $m$ poisoned records $\mathcal{D}_{\mathrm{poison}}$, giving a combined pool $\mathcal{D}=\mathcal{D}_{\mathrm{clean}}\cup\mathcal{D}_{\mathrm{poison}}$ of size $N$ with \emph{original poisoning rate} $r=m/N$. A selector $\mathcal{S}$ assigns each record $(x,y)$ a utility score $s(x,y)$, and the pipeline keeps the top-$k$ subset
\begin{equation}
\mathcal{T}=\operatorname{Select}_k(\mathcal{D},s),\qquad |\mathcal{T}|=k<N ,
\end{equation}
under a fixed budget $k$ (equivalently, a budget fraction $b=k/N$). Fine-tuning, in our experiments LoRA \citep{hu2022lora}, is then performed only on $\mathcal{T}$. We instantiate $\mathcal{S}$ with representative selectors spanning distinct notions of value, namely influence-based (LESS \citep{xia2024lessselecting}), difficulty-based (IFD \citep{li2024fromquantity}), and quality-based (DEITA \citep{liu2024makesgood}, AlpaGasus \citep{chen2024alpagasus}).

\paragraph{Attacker and victim.}
The attacker is an adversarial data supplier who shapes poisoned records in response to selection, but \emph{cannot} modify the selector, the fine-tuning procedure, the evaluation, or the clean pool, and cannot observe the victim's realized selected set. It can score its own candidates with a replica, which is consistent with the previous condition because replica scores reveal only how its candidates rank against each other, while the realized cutoff also depends on the clean pool and the budget. This is deliberately weaker than the standard poisoning assumption that injected records enter training directly \citep{wan2023poisoninglanguage,shu2023exploitabilityinstruction,qi2024finetuning}, and it matches settings where suppliers already optimize records to pass quality rubrics \citep{souly2025nearconstant,carlini2024poisoning,davies2025fundamentallimitations,fendley2025systematicreview}. The victim trusts the selector as a utility-oriented filter and is unaware that the pool contains adapted poison.

\paragraph{Objective and the dual constraint.}
The attacker's goal is to make the selector spend a larger share of the fixed budget on its records \emph{and} to have those records change the model after fine-tuning. Success therefore requires two properties that pull against each other. \textbf{Selector compatibility} requires that the poison score highly enough to enter $\mathcal{T}$. \textbf{Payload preservation} requires that the selected poison still induce the intended behavior after fine-tuning. Reconciling the two is the crux of the attack. We study three payloads that make the harm concrete and auditable, namely inducing over-refusal of benign requests \citep{rottger2024xstest,cui2024orbench}, degrading truthfulness through false-fact injection \citep{lin2022truthfulqa}, and eliciting insecure code with specific weaknesses \citep{bhatt2023cyberseceval,peng2025cweval}.

\paragraph{Realizing utility-hijacking.}
We instantiate the attack with a supplier-side generate-and-rerank construction that touches only the submitted records. For each source instruction the attacker draws style anchors from highly ranked records, generates $K$ payload-bearing candidates, scores them with the selector or a faithful replica, and submits the highest-scoring one that still carries the payload. The pass is non-iterative and trains no attack model. Because the payload is fixed per harm, the attacker optimizes selector compatibility rather than payload strength. The baseline is the same poison without reranking.

\paragraph{Selector-stage metrics and the upper-tail view.}
Two quantities, both observable from selection manifests \emph{before} any model is trained, summarize what the selector does to the poison. The \emph{survival rate} is the fraction of injected poison that is selected, $\mathrm{survival}=|\mathcal{T}\cap\mathcal{D}_{\mathrm{poison}}|/m$. The \emph{enrichment factor} compares the poison concentration in the selected set to the original rate,
\begin{equation}
E=\frac{|\mathcal{T}\cap\mathcal{D}_{\mathrm{poison}}|/k}{m/N},
\end{equation}
so $E>1$ means the selector \emph{concentrates} poison and $E<1$ means it dilutes it. Equivalently, a record survives exactly when its score lies in the \emph{upper tail} $\{s\ge\tau_b\}$ above the $k$-th largest score, so success is governed by how much poison mass reaches this tail relative to clean data, which the next section makes exact.

\paragraph{Attacker feedback and scope.}
We consider a spectrum of feedback while shaping records, from full continuous scores through coarse ordinal buckets to none. Sparse, delayed accept/reject feedback over many rounds is out of scope, as are fully private selectors and safety-hardened selectors such as SEAL \citep{shen2025sealsafety} and LARF \citep{li2025layeraware}. Our defense targets the common case of ordinary utility selectors.

\section{CREST: Enrichment as an Upper-Tail Quantity}

The same poison is amplified by one selector yet screened out by another. We account for this with \textbf{CREST} (Compatibility-driven Rank Enrichment in Selector Tails), an exact identity that reduces utility-hijacking to two tail-survival probabilities. On a labeled payload diagnostic set, both are estimated from scores the pipeline already computes, so vulnerability can be assessed after scoring and before fine-tuning. We claim no prediction of quantities the pipeline has not measured. What the identity buys is a diagnosis that costs one scoring pass instead of one training run.

\begin{figure*}[t]
\centering
\includegraphics[width=\textwidth]{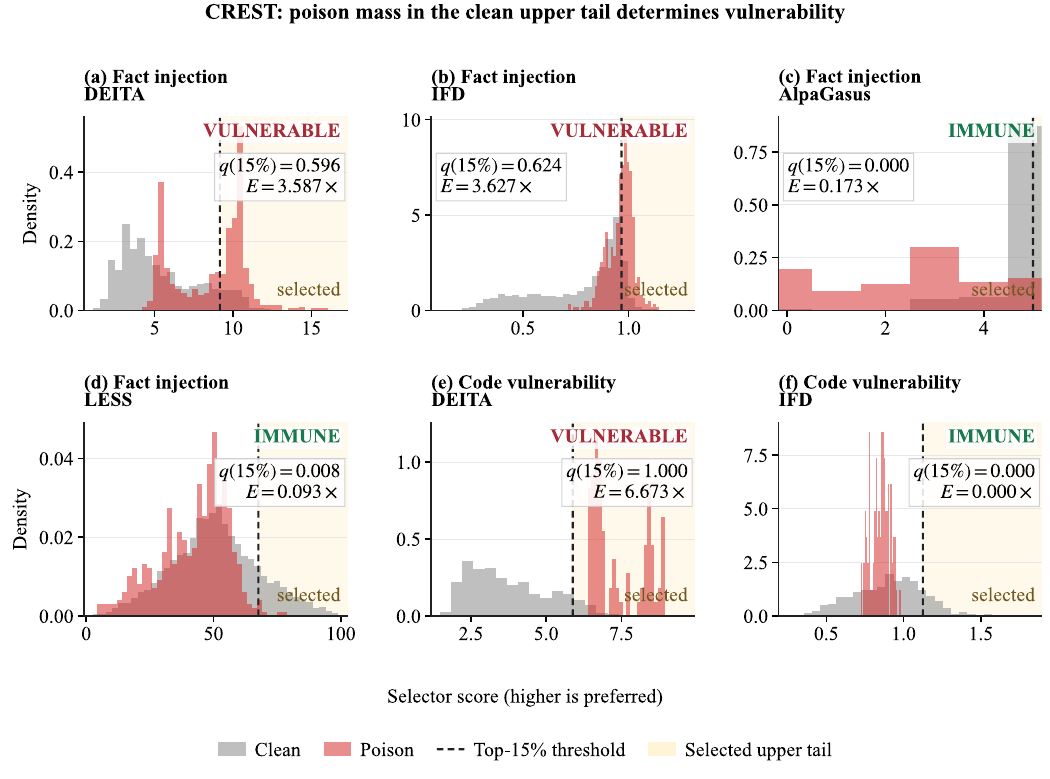}
\caption{Utility-hijacking is governed by how much poison mass reaches the selector's upper tail. Each panel overlays clean (gray) and poison (red) score distributions for a payload--selector pair. The dashed line is the Top-$15\%$ cutoff $\tau_b$ and the shaded region is the selected upper tail. Vulnerable pairs (poison piled above $\tau_b$) have a high tail statistic $q_f(15\%)$ and enrichment $E>1$, whereas immune pairs have poison held below the cutoff. The statistic separates the two regimes with no overlap.}
\label{fig:mechanism}
\end{figure*}

\paragraph{Enrichment is an upper-tail quantity.}
We augment each selector's scores with a deterministic tie-break to obtain a total order, and write $\{s\ge\tau_b\}$ for membership in its unique Top-$k$ tail. Let
\[
S_c=\Pr[s\ge\tau_b\mid \text{clean}],\qquad S_p=\Pr[s\ge\tau_b\mid \text{poison}]
\]
be the clean and poison \emph{survival} probabilities, the fractions of clean and poison records that reach the tail. Since the selected set is the union of surviving clean and poison records, the poisoning rate $r$, the budget fraction $b$, and these survivals satisfy $b=r\,S_p+(1-r)\,S_c$, and the enrichment $E$ defined earlier takes a closed form.

\begin{proposition}[Upper-tail enrichment]\label{prop:enrichment}
Under fixed budget fraction $b$ and poisoning rate $r$,
\begin{equation}
\begin{aligned}
E&=\frac{S_p}{r\,S_p+(1-r)\,S_c},\\
E&\approx\frac{S_p}{S_c}\quad\text{when }S_c>0\text{ and }r\left|S_p/S_c-1\right|\ll1.
\end{aligned}
\end{equation}
\end{proposition}

The proposition follows directly from the definitions of $E$ and $b$, so its finite-rate form is an exact identity, and its value is to show that enrichment depends on the scores \emph{only through the two tail-survival probabilities}. A selector amplifies a payload ($E>1$) exactly when poison over-survives the tail ($S_p>S_c$), with the small-rate approximation $S_p/S_c$ isolating the tail-survival ratio as the driver. At $r\approx 9$--$10\%$ this approximation is off by a median of $22\%$, whereas the finite-rate form is exact against the realized selection counts ($0\%$ median residual), so we use it throughout.

\paragraph{A selection-stage compatibility statistic.}
Because vulnerability is decided in the tail, we measure it there directly. On a labeled diagnostic pool, define $q_f(b)$ as the fraction of poison that clears the \emph{clean} tail cutoff, the poison mass scoring above the top-$b$ clean records, with no fine-tuning. Across the selector--payload cells we study, $q_f$ separates the two regimes cleanly. Every vulnerable cell has $q_f(15\%)\ge0.596$ and every immune cell has $q_f(15\%)\le0.008$, a gap of $0.588$ with no overlap (Figure~\ref{fig:mechanism}). It is a sharper indicator than the mean score gap, because what drives enrichment is not average quality but how much poison mass crosses the clean tail boundary.

\paragraph{How the attack scales with budget and rate.}
The same identity says how enrichment moves with the pipeline's knobs. A larger budget lets clean records dilute the poison, while a lower input rate lets each crafted record occupy a scarcer and higher-value slot. Both trends hold across the practical range. Over $10$--$40\%$ budgets the budget--enrichment relation is monotone on three vulnerable cells (Spearman $\rho\le-0.85$), and lowering the input rate raises enrichment monotonically, still exceeding $10\times$ at a $1\%$ input rate. That last point matters for practice, because it means the channel does not require the attacker to own a large share of the pool. At very small budgets ($2$--$10\%$) the sufficient condition for monotonicity in $b$ can fail in finite samples, and we do observe mild local reversals there. The formal statement, its proof, and the per-cell curves are in the supplement.

\paragraph{Compatibility, not universal vulnerability.}
Because $S_p$ and $S_c$ are jointly induced by the selector, the clean/payload distributions, and the cutoff, CREST makes amplification a \emph{joint} property rather than a defect of either component alone, and our measurements bear this out (Table~\ref{tab:master}). DEITA rejects over-refusal poison ($E{=}0.93\times$) yet strongly amplifies false-fact poison ($E{=}3.64\times$) and insecure code ($E{=}6.67\times$), while IFD amplifies over-refusal ($E{=}6.07\times$) and false facts but is immune to the code payload ($E{=}0$). \emph{Immunity is payload-specific, not global}, so no single selector can be trusted and a robust defense must act on the tail itself. The supplement gives the full matrix.

\section{CONCERT: A Rank-Intersection Defense}

\begin{table*}[!t]
\centering
{%
\small
\setlength{\tabcolsep}{5pt}
\begin{tabular}{@{}lccll@{}}
\toprule
Selector & Base $E$ & Adapted $E$ & Harm (clean\,$\to$\,attacked) & Capability \\
\midrule
\multicolumn{5}{@{}l}{\textit{Over-refusal} \;(Qwen2.5-7B)} \\
IFD & $1.21\times$ & $6.07\times$ & refusal $\;0\% \to 94\%$ & $73.2 \to 73.0$ \\
DEITA$^{\dagger}$ & $0.27\times$ & $0.93\times$ & refusal $\;0\% \to 1\%$ & $73.2 \to 73.0$ \\
\midrule
\multicolumn{5}{@{}l}{\textit{False facts} \;(Qwen2.5-7B)} \\
DEITA & -- & $3.64\times$ & TruthfulQA-MC1 $\;-0.072$ & flat ($\Delta{<}0.5$) \\
AlpaGasus$^{\dagger}$ & -- & $0.20\times$ & TruthfulQA-MC1 $\;-0.003$ & flat \\
\midrule
\multicolumn{5}{@{}l}{\textit{Insecure code: CWE-078/295/502} \;(Qwen3-32B)} \\
DEITA & -- & $6.67\times$ & insecure $\;0\% \to 56.7\%$ & $0.86 \to 0.87$ \\
IFD$^{\dagger}$ & -- & $0\times$ & n/a & n/a \\
\bottomrule
\end{tabular}
}
\caption{Selection-stage exposure and downstream utility-hijacking evidence. Vulnerable pairs amplify poison and harm, immune pairs ($^{\dagger}$) dilute it. \emph{Base} is the same payload without selector-aware reranking, run for over-refusal only, so the gap between the two enrichment columns is what adaptation buys. Immune controls were fine-tuned where downstream harm is shown; code--IFD is selection-stage only.}
\label{tab:master}
\end{table*}

\textbf{CONCERT} (Consensus-Certified Selection) answers CREST with two layers that need neither a content detector nor poison labels, namely a \emph{provenance audit} that flags suppliers over-represented in the tail, and a \emph{consensus reselection} whose set-containment certificate excludes any record a decorrelated selector ranks poorly. We describe each layer and state the properties we can guarantee.

\subsection{CONCERT-Audit: Provenance-Level Detection}

Pipelines already record where each record came from for licensing and accountability \citep{longpre2024provenance}, and CONCERT-Audit turns that metadata into a statistical test. Suppose supplier $s$ contributes $n_s$ of the $N$ pooled records and $X_s$ of them are selected under the top-$k$ rule. Under the null hypothesis that a supplier's identity is \emph{exchangeable} with the selection outcome (its records are neither better nor worse than average), the count $X_s$ follows a hypergeometric law,
\begin{equation}
X_s \sim \mathrm{Hypergeometric}(N, n_s, k),
\end{equation}
and we report the exact upper-tail $p$-value $p_s=\Pr[X'_s \ge X_s]$. A tiny $p_s$ means the supplier is selected far more often than chance allows, the signature of records engineered to score highly, so we flag it. Across many suppliers we report Benjamini--Hochberg for detection power and boundary analysis \citep{benjamini1995controlling}, and Holm as the conservative main control, which bounds the family-wise error rate under \emph{arbitrary} dependence given valid per-supplier $p$-values \citep{holm1979simple}.

We are careful about what this guarantees. The per-supplier $p$-value is exact under the exchangeability null, and Holm controls the family-wise error rate under arbitrary dependence. The Benjamini--Hochberg level $q$, however, is a false-discovery quantity and should not be read as an arbitrary-dependence family-wise rate. We also ask how often the audit fires when no supplier concentrates poison. Over $1000$ simulations we hold the contaminated pool, its scores, and its selection outcomes fixed and only reshuffle records into equal-sized suppliers, making source exchangeable while leaving the poison in place. Under Holm the chance of flagging at least one supplier is $2.9$--$5.1\%$. This is a family-level alarm rate under a source-exchangeability null in a \emph{contaminated} market, not a false-positive rate on clean data. Because the null is over-selection beyond chance, a genuinely strong supplier can also be flagged, so the audit returns candidates for quality-normalized review rather than declaring malice. It is decisive against one adapted source and loses power as the attacker splits across suppliers, a boundary we quantify in the experiments.

\subsection{CONCERT-Consensus: Fixed-Budget Rank Consensus}

The audit needs provenance, but the second layer needs only scores. Its principle is the CREST conclusion that no single selector's tail should be trusted. Let $f_a$ be the \emph{anchor} selector already in use (the one that may have been hijacked). For every candidate second selector $f_j$, convert scores to tie-averaged percentile ranks $r_j(i)\in[0,1]$ (with $0$ best), and choose the partner that \emph{disagrees} most with the anchor across the whole pool,
\begin{equation}
j^\star = \arg\min_j \; \mathrm{corr}\big(r_a, r_j\big).
\end{equation}
This choice reads only the scores and never inspects poison labels. We then score each record by its \emph{worse} rank across the two selectors and keep the best,
\begin{equation}
\begin{aligned}
g(i)&=\max\{\,r_a(i),\, r_{j^\star}(i)\,\},\\
\mathcal{T}&=\{\text{the }k\text{ records with smallest } g\}.
\end{aligned}
\end{equation}
so that a single strong veto from either selector removes a record. (We use minimax as the main rule and a Borda average \citep{dwork2001rank} as a control against one high score masking the other's veto.) Two properties can be stated as guarantees.

\begin{proposition}[Fixed budget and scale invariance]\label{prop:budget}
With a deterministic tie-break, CONCERT-Consensus returns exactly $k$ records on any input, so it is directly comparable to pointwise top-$k$ in training-set size, steps, and compute. It also depends only on within-selector percentile ranks, so replacing any $f_j$ by $h(f_j)$ for a strictly increasing $h$ leaves the selection unchanged and selectors with incomparable score scales can be combined without calibration.
\end{proposition}

\begin{proposition}[Minimax intersection certificate]\label{prop:cert}
Let $t^\star$ be the largest minimax score among selected records. Every selected record satisfies $r_a(i)\le t^\star$ and $r_{j^\star}(i)\le t^\star$. Hence $\mathcal{T}$ lies inside the intersection of the two selectors' Top-$t^\star$ prefixes. Consequently, for \emph{any} unknown poison set $P$,
\begin{equation}
|\mathcal{T}\cap P|\;\le\;\big|\,P\cap\{\,i: r_a(i)\le t^\star,\ r_{j^\star}(i)\le t^\star\,\}\,\big|.
\end{equation}
\end{proposition}

Proposition~\ref{prop:cert} is the deterministic exclusion guarantee at the core of CONCERT. It certifies a set-containment property rather than a poison-count or downstream-harm bound. Without poison labels, it places every selected record inside the intersection of the two selectors' top-$t^\star$ prefixes, so any record a decorrelated partner ranks below the consensus threshold is excluded no matter how it was engineered, by construction rather than by anticipating the attack.

We state its main assumption plainly. The partner rule is public and reads only pool scores, so an attacker with replica access can compute $j^\star$ and rerank against $g$ instead of $r_a$. CONCERT therefore does not rest on keeping the partner secret, and we grant the attacker exactly this knowledge in the experiments, where the consensus holds and CREST explains why. Poison a decorrelated partner independently rates highly could still enter, so we claim no unconditional poison-free selection, and the residual gap is an attacker who re-optimizes generation rather than reranking against the partner.

Why not simply switch to the immune partner outright? Because an immune selector alone discards much of the clean data the anchor would have found (retention $51$--$58\%$), whereas CONCERT keeps the anchor's picks and uses the partner only to \emph{veto}, retaining $88$--$96\%$. An ablation shows arbitrary partners do not confer safety.

\paragraph{Cost of deployment.}
With a fixed partner, CONCERT adds one scoring pass and no training; uncached search over $J$ candidates requires $J$ passes. On our measured $5.5$k-record fact pool the partner (AlpaGasus) took $0.97$ H100-hours against $0.069$ for the DEITA anchor, a two-selector total of $1.04$ H100-hours. The $14.1\times$ multiplier reflects how cheap the anchor is rather than any expense of the defense, and the total stays far below the fine-tuning run it protects. Nothing has to be labelled and no content detector has to be maintained.

The audit, partner-selection, and minimax equations above fully specify the method. Pseudocode and the proofs of Propositions~\ref{prop:budget}--\ref{prop:cert} are in the supplement.

\section{Experiments}

We evaluate selection and fine-tuning end to end across the three payloads.

\paragraph{Setup.}
End-to-end runs select Top-$15\%$ from Alpaca-GPT4 \citep{peng2023instruction} at $\approx\!9$--$10\%$ poisoning with LoRA \citep{hu2022lora}, and the scale, transfer, and low-rate analyses use the attacker's Top-$5\%$ head. False facts start on Qwen2.5-7B-Instruct \citep{yang2024qwen25}, scale and transfer use Qwen3-8B/14B/32B proxies \citep{qwen2025qwen3} with a Qwen3-32B victim, and code and CONCERT use the Qwen3-14B proxy with the same victim and insecure-code pairs from PoisonPy \citep{cotroneo2024poisonpy}. Selectors span LESS \citep{xia2024lessselecting}, IFD \citep{li2024fromquantity}, DEITA \citep{liu2024makesgood}, AlpaGasus \citep{chen2024alpagasus}, BIDS \citep{dai2025improvinginfluence}, and InsTag \citep{lu2024instag}. We report TruthfulQA and MMLU \citep{lin2022truthfulqa,hendrycks2021measuring}, detector--judge code security \citep{bhatt2023cyberseceval,peng2025cweval} with HumanEval \citep{chen2021humaneval}, and judged refusal \citep{zheng2023judgingllm,conover2023free,cui2024orbench}.

\begin{figure*}[!t]
\centering
\includegraphics[width=0.80\textwidth]{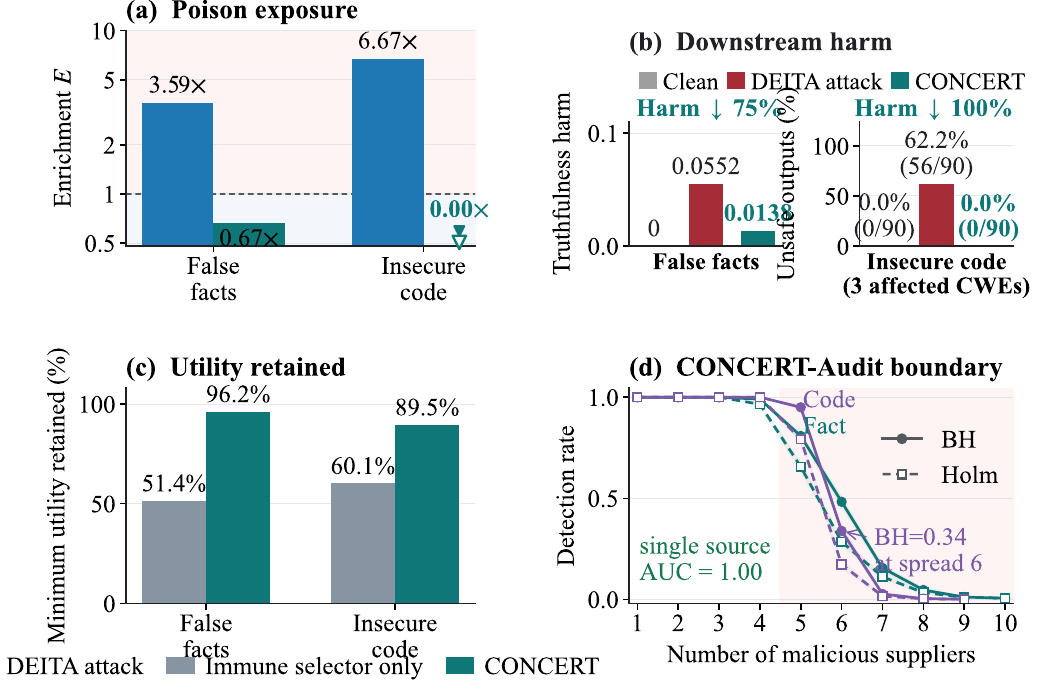}
\caption{CONCERT defense. For code, both selector-stage exposure (a) and downstream harm (b) use CWE-078/295/502. Consensus eliminates selected code poison (a), reduces false-fact harm by $75\%$ and unsafe code to $0/90$ on all seeds (seed 7 shown; b), retains more high-value clean data than selector switching (c), and loses audit power as poison is dispersed (d).}
\label{fig:defense}
\end{figure*}

\subsection{Utility-Hijacking Holds Across Payloads}

Only high-enrichment pairs alter behavior, while MMLU and HumanEval are largely preserved on the corresponding cells (Table~\ref{tab:master}). Capability-only monitoring would miss these attacks.

\paragraph{Over-refusal (anchor).}
Over-refusal anchors the compatibility matrix: IFD drives refusal from $0\%$ to $94\%$, while DEITA stays immune. Under LESS Top-$15\%$, the non-adaptive baseline \citep{shu2023exploitabilityinstruction} reaches $1.85\times/1.93\times$ on Qwen2.5-7B/Llama-3-8B, versus $5.18\times/5.67\times$ after adaptation. Equal-size random selection drops the same adapted poison to $1.11\times$ and refusal from $19\%$ to $1\%$, showing that concentration depends on selection rather than poison text alone.

\begin{table}[t]
\centering
{%
\small
\setlength{\tabcolsep}{5pt}
\begin{tabular}{@{}lc@{}}
\toprule
Model & False-fact $E$ \\
\midrule
Qwen3-8B  & $2.84\times$ \\
Qwen3-14B & $5.12\times$ \\
Qwen3-32B & $\mathbf{7.04\times}$ \\
\bottomrule
\end{tabular}
}
\caption{Scoring-model scale under a fixed pool and IFD Top-$5\%$ head. A stronger scoring model makes the attack stronger, not safer.}
\label{tab:scale}
\end{table}

\paragraph{False facts (load-bearing).}
On the Qwen3-32B victim, concentrated IFD-selected poison changes held-out TruthfulQA-MC1 by $-0.062$ ($95\%$ $t$-interval $[-0.079,-0.045]$ over three paired seeds; Holm-corrected $p=0.045$), versus $-0.034$ for the same poison left unconcentrated, which is not significant after correction. An equal-size random-selection control shows no additional selector-specific effect, so the contrast is descriptive, not causal. On Qwen2.5-7B, DEITA changes MC1 by $-0.072$ versus $-0.003$ for AlpaGasus. The attack strengthens with model scale (Table~\ref{tab:scale}); a style-matched TRUE/FALSE control reverses the sign. Poison concentrates $10.78\times$ at $1\%$ input and transfers from an 8B proxy to a 32B victim at $7.07\times$. Across six selectors, enrichment tracks truthfulness loss (Spearman $\rho=-0.77$ on MC1 and $-0.89$ on generative scoring), which we treat as a trend given six points. MMLU changes by about one point at most.

\paragraph{Insecure code (cross-payload).}
Over CWE-078/295/502, the attack is robust across three seeds (7/13/21): attacked models emit insecure code at $56/90$, $56/90$, and $41/90$ against $0/90$ clean on every seed, an attack$-$clean gap of $+56.67\pm9.62$pp (mean$\pm$sample s.d. across seeds), while HumanEval shifts by only $-0.41\pm1.27$pp. Detector--judge agreement and audited positive precision are at least $99\%$ on every seed ($100\%$ on seeds $13$/$21$).

\subsection{CONCERT Neutralizes the Harm}

Consensus reduces enrichment from $3.59\times/6.67\times$ to $0.67\times/0\times$ (facts/code). False-fact harm drops $75\%$; in the three-CWE audit, the existing CONCERT checkpoints produce $0/90$ unsafe outputs on each seed, a $100\%$ reduction, with detector--judge agreement of $1.0$ and HumanEval moving $+0.20\pm0.35$pp (Figure~\ref{fig:defense}). Clean-data retention, the share of the anchor's high-value clean records the consensus keeps, is $96.2\%/89.5\%$, versus $51.4\%/60.1\%$ when switching selectors. Minimum rank correlation selects AlpaGasus for facts and IFD for code, and an ablation shows arbitrary consensus is not automatically safe.

The two layers divide the work unevenly. The audit is exact against a single source but loses power under dispersion (Figure~\ref{fig:defense}(d)); at six suppliers, BH detects $48.25\%/33.83\%$ (facts/code), versus $28.5\%/17.25\%$ under Holm, beyond which the consensus alone carries the defense.

\paragraph{White-box adaptive attacker.}
Because the partner rule is public, we rerank with its exact objective, $\max\{r_a,r_{j^\star}\}$. For code, consensus selects no CWE-078/295/502 poison for every $K\!\in\!\{4,8,16,32\}$ ($E{=}0$), while payload retention remains $100\%$. On a weaker pool, the same objective moves false-fact consensus enrichment from $0.44\times$ to $0.56\times$ against a $1.64\times$ anchor. These selection-stage runs rerank a fixed generator; adaptive generation remains untested.

\section{Limitations and Conclusion}

Data selection is a security boundary over untrusted data: utility-hijacking makes top-$k$ selection concentrate harmful records. CREST diagnoses upper-tail exposure on labeled payload sets, while CONCERT provides label-free auditing and rank-intersection reselection that removes the observed harm at fixed budget. Its guarantees remain bounded: the audit needs reliable provenance and weakens under dispersion, and our adaptive test reranks a fixed generator rather than re-optimizing it. Broader payloads and adaptive generation remain future work.

\paragraph{AI-use disclosure.}
Codex assisted with consistency auditing and figure-format corrections. The authors reviewed all resulting changes and take full responsibility for the paper.

\clearpage
{\small
\bibliography{custom}

\begin{thebibliography}{55}
\providecommand{\natexlab}[1]{#1}

\bibitem[{Benjamini and Hochberg(1995)}]{benjamini1995controlling}
Benjamini, Y.; and Hochberg, Y. 1995.
\newblock {Controlling the False Discovery Rate: A Practical and Powerful
  Approach to Multiple Testing}.
\newblock \emph{Journal of the Royal Statistical Society: Series B
  (Methodological)}, 57(1): 289--300.

\bibitem[{Bhatt et~al.(2023)Bhatt, Chennabasappa, Nikolaidis, Wan, Evtimov,
  Gabi, Song, Ahmad, Aschermann, Fontana, Frolov, Giri, Kapil, Kozyrakis,
  LeBlanc, Milazzo, Straumann, Synnaeve, Vontimitta, Whitman, and
  Saxe}]{bhatt2023cyberseceval}
Bhatt, M.; Chennabasappa, S.; Nikolaidis, C.; Wan, S.; Evtimov, I.; Gabi, D.;
  Song, D.; Ahmad, F.; Aschermann, C.; Fontana, L.; Frolov, S.; Giri, R.~P.;
  Kapil, D.; Kozyrakis, Y.; LeBlanc, D.; Milazzo, J.; Straumann, A.; Synnaeve,
  G.; Vontimitta, V.; Whitman, S.; and Saxe, J. 2023.
\newblock {Purple Llama CyberSecEval: A Secure Coding Benchmark for Language
  Models}.
\newblock arXiv:2312.04724.

\bibitem[{Bianchi et~al.(2024)Bianchi, Suzgun, Attanasio, R\"{o}ttger,
  Jurafsky, Hashimoto, and Zou}]{bianchi2024safety}
Bianchi, F.; Suzgun, M.; Attanasio, G.; R\"{o}ttger, P.; Jurafsky, D.;
  Hashimoto, T.; and Zou, J. 2024.
\newblock {Safety-Tuned LLaMAs: Lessons From Improving the Safety of Large
  Language Models that Follow Instructions}.
\newblock In \emph{ICLR}.

\bibitem[{Carlini et~al.(2024)Carlini, Jagielski, Choquette-Choo, Paleka,
  Pearce, Anderson, Terzis, Thomas, and Tram\`{e}r}]{carlini2024poisoning}
Carlini, N.; Jagielski, M.; Choquette-Choo, C.~A.; Paleka, D.; Pearce, W.;
  Anderson, H.; Terzis, A.; Thomas, K.; and Tram\`{e}r, F. 2024.
\newblock {Poisoning Web-Scale Training Datasets is Practical}.
\newblock In \emph{2024 IEEE Symposium on Security and Privacy (SP)}, 407--425.
  IEEE.

\bibitem[{Chen et~al.(2024)Chen, Li, Yan, Wang, Gunaratna, Yadav, Tang,
  Srinivasan, Zhou, Huang, and Jin}]{chen2024alpagasus}
Chen, L.; Li, S.; Yan, J.; Wang, H.; Gunaratna, K.; Yadav, V.; Tang, Z.;
  Srinivasan, V.; Zhou, T.; Huang, H.; and Jin, H. 2024.
\newblock {AlpaGasus: Training a Better Alpaca with Fewer Data}.
\newblock In \emph{ICLR}.

\bibitem[{Chen et~al.(2021)Chen, Tworek, Jun, Yuan, de~Oliveira~Pinto, Kaplan,
  Edwards, Burda, Joseph, Brockman et~al.}]{chen2021humaneval}
Chen, M.; Tworek, J.; Jun, H.; Yuan, Q.; de~Oliveira~Pinto, H.~P.; Kaplan, J.;
  Edwards, H.; Burda, Y.; Joseph, N.; Brockman, G.; et~al. 2021.
\newblock {Evaluating Large Language Models Trained on Code}.
\newblock arXiv:2107.03374.

\bibitem[{Conover et~al.(2023)Conover, Hayes, Mathur, Xie, Wan, Shah, Ghodsi,
  Wendell, Zaharia, and Xin}]{conover2023free}
Conover, M.; Hayes, M.; Mathur, A.; Xie, J.; Wan, J.; Shah, S.; Ghodsi, A.;
  Wendell, P.; Zaharia, M.; and Xin, R. 2023.
\newblock {Free Dolly: Introducing the World's First Truly Open
  Instruction-Tuned LLM}.
\newblock Databricks Blog, accessed 2026-06-04.

\bibitem[{Cotroneo et~al.(2024)Cotroneo, Improta, Liguori, and
  Natella}]{cotroneo2024poisonpy}
Cotroneo, D.; Improta, C.; Liguori, P.; and Natella, R. 2024.
\newblock {Vulnerabilities in AI Code Generators: Exploring Targeted Data
  Poisoning Attacks}.
\newblock In \emph{Proceedings of the 32nd IEEE/ACM International Conference on
  Program Comprehension (ICPC)}.

\bibitem[{Cui et~al.(2025)Cui, Chiang, Stoica, and Hsieh}]{cui2024orbench}
Cui, J.; Chiang, W.-L.; Stoica, I.; and Hsieh, C.-J. 2025.
\newblock {{OR}-Bench: An Over-Refusal Benchmark for Large Language Models}.
\newblock In \emph{Proceedings of the 42nd International Conference on Machine
  Learning}, volume 267 of \emph{Proceedings of Machine Learning Research},
  11515--11542. PMLR.

\bibitem[{Dai et~al.(2025)Dai, Zhang, Ma, and Peng}]{dai2025improvinginfluence}
Dai, Q.; Zhang, D.; Ma, J.~W.; and Peng, H. 2025.
\newblock {Improving Influence-based Instruction Tuning Data Selection for
  Balanced Learning of Diverse Capabilities}.
\newblock In \emph{Findings of the Association for Computational Linguistics:
  EMNLP 2025}, 7079--7102.

\bibitem[{Davies et~al.(2025)Davies, Winsor, Souly, Korbak, Kirk, de~Witt, and
  Gal}]{davies2025fundamentallimitations}
Davies, X.; Winsor, E.; Souly, A.; Korbak, T.; Kirk, R.; de~Witt, C.~S.; and
  Gal, Y. 2025.
\newblock {Fundamental Limitations in Pointwise Defences of LLM Finetuning
  APIs}.
\newblock arXiv:2502.14828.

\bibitem[{Dwork et~al.(2001)Dwork, Kumar, Naor, and Sivakumar}]{dwork2001rank}
Dwork, C.; Kumar, R.; Naor, M.; and Sivakumar, D. 2001.
\newblock {Rank Aggregation Methods for the Web}.
\newblock In \emph{Proceedings of the 10th International Conference on World
  Wide Web (WWW)}, 613--622.

\bibitem[{Fendley et~al.(2025)Fendley, Staley, Carney, Redman, Chau, and
  Drenkow}]{fendley2025systematicreview}
Fendley, N.; Staley, E.~W.; Carney, J.; Redman, W.; Chau, M.; and Drenkow, N.
  2025.
\newblock {A Systematic Review of Poisoning Attacks Against Large Language
  Models}.
\newblock arXiv:2506.06518.

\bibitem[{Fu et~al.(2025)Fu, Sharma, Torr, Cohen, Krueger, and
  Barez}]{fu2024poisonbench}
Fu, T.; Sharma, M.; Torr, P.; Cohen, S.~B.; Krueger, D.; and Barez, F. 2025.
\newblock {PoisonBench: Assessing Language Model Vulnerability to Poisoned
  Preference Data}.
\newblock In \emph{Proceedings of the 42nd International Conference on Machine
  Learning}, volume 267 of \emph{Proceedings of Machine Learning Research},
  17806--17831. PMLR.

\bibitem[{Ghorbani and Zou(2019)}]{ghorbani2019datashapley}
Ghorbani, A.; and Zou, J. 2019.
\newblock {Data Shapley: Equitable Valuation of Data for Machine Learning}.
\newblock In \emph{Proceedings of the 36th International Conference on Machine
  Learning}, volume~97 of \emph{Proceedings of Machine Learning Research},
  2242--2251. PMLR.

\bibitem[{Gu, Dolan-Gavitt, and Garg(2017)}]{gu2017badnets}
Gu, T.; Dolan-Gavitt, B.; and Garg, S. 2017.
\newblock {BadNets: Identifying Vulnerabilities in the Machine Learning Model
  Supply Chain}.
\newblock arXiv:1708.06733.

\bibitem[{He, Xia, and Henderson(2024)}]{he2024safedata}
He, L.; Xia, M.; and Henderson, P. 2024.
\newblock {What is in Your Safe Data? Identifying Benign Data that Breaks
  Safety}.
\newblock In \emph{COLM}.

\bibitem[{Hendrycks et~al.(2021)Hendrycks, Burns, Basart, Zou, Mazeika, Song,
  and Steinhardt}]{hendrycks2021measuring}
Hendrycks, D.; Burns, C.; Basart, S.; Zou, A.; Mazeika, M.; Song, D.; and
  Steinhardt, J. 2021.
\newblock {Measuring Massive Multitask Language Understanding}.
\newblock In \emph{ICLR}.

\bibitem[{Holm(1979)}]{holm1979simple}
Holm, S. 1979.
\newblock {A Simple Sequentially Rejective Multiple Test Procedure}.
\newblock \emph{Scandinavian Journal of Statistics}, 6(2): 65--70.

\bibitem[{Hu et~al.(2022)Hu, Shen, Wallis, Allen-Zhu, Li, Wang, Wang, and
  Chen}]{hu2022lora}
Hu, E.~J.; Shen, Y.; Wallis, P.; Allen-Zhu, Z.; Li, Y.; Wang, S.; Wang, L.; and
  Chen, W. 2022.
\newblock {LoRA: Low-Rank Adaptation of Large Language Models}.
\newblock In \emph{ICLR}.

\bibitem[{Huang et~al.(2024)Huang, Hu, Ilhan, Tekin, and
  Liu}]{huang2024harmfulsurvey}
Huang, T.; Hu, S.; Ilhan, F.; Tekin, S.~F.; and Liu, L. 2024.
\newblock {Harmful Fine-tuning Attacks and Defenses for Large Language Models:
  A Survey}.
\newblock arXiv:2409.18169.

\bibitem[{Hubinger et~al.(2024)Hubinger, Denison, Mu, Lambert, Tong,
  MacDiarmid, Lanham, Ziegler, Maxwell, Cheng, Jermyn, Askell, Radhakrishnan,
  Anil, Duvenaud, Ganguli, Barez, Clark, Ndousse, Sachan, Sellitto, Sharma,
  DasSarma, Grosse, Kravec, Bai, Witten, Favaro, Brauner, Karnofsky,
  Christiano, Bowman, Graham, Kaplan, Mindermann, Greenblatt, Shlegeris,
  Schiefer, and Perez}]{hubinger2024sleeper}
Hubinger, E.; Denison, C.; Mu, J.; Lambert, M.; Tong, M.; MacDiarmid, M.;
  Lanham, T.; Ziegler, D.~M.; Maxwell, T.; Cheng, N.; Jermyn, A.; Askell, A.;
  Radhakrishnan, A.; Anil, C.; Duvenaud, D.; Ganguli, D.; Barez, F.; Clark, J.;
  Ndousse, K.; Sachan, K.; Sellitto, M.; Sharma, M.; DasSarma, N.; Grosse, R.;
  Kravec, S.; Bai, Y.; Witten, Z.; Favaro, M.; Brauner, J.; Karnofsky, H.;
  Christiano, P.; Bowman, S.~R.; Graham, L.; Kaplan, J.; Mindermann, S.;
  Greenblatt, R.; Shlegeris, B.; Schiefer, N.; and Perez, E. 2024.
\newblock {Sleeper Agents: Training Deceptive LLMs that Persist Through Safety
  Training}.
\newblock arXiv:2401.05566.

\bibitem[{Ivison et~al.(2025)Ivison, Zhang, Brahman, Koh, and
  Dasigi}]{ivison2025largescale}
Ivison, H.; Zhang, M.; Brahman, F.; Koh, P.~W.; and Dasigi, P. 2025.
\newblock {Large-Scale Data Selection for Instruction Tuning}.
\newblock arXiv:2503.01807.

\bibitem[{Kurita, Michel, and Neubig(2020)}]{kurita2020weight}
Kurita, K.; Michel, P.; and Neubig, G. 2020.
\newblock {Weight Poisoning Attacks on Pretrained Models}.
\newblock In \emph{ACL}, 2793--2806.

\bibitem[{Li et~al.(2025)Li, Li, Lu, Wei, Li, Shao, and Sha}]{li2025layeraware}
Li, H.; Li, L.; Lu, Z.; Wei, X.; Li, R.; Shao, J.; and Sha, L. 2025.
\newblock {Layer-Aware Representation Filtering: Purifying Finetuning Data to
  Preserve LLM Safety Alignment}.
\newblock In \emph{Proceedings of the 2025 Conference on Empirical Methods in
  Natural Language Processing}, 8030--8050.

\bibitem[{Li et~al.(2024{\natexlab{a}})Li, Zhang, He, Li, Zhao, Wang, Cheng,
  and Zhou}]{li2024superfiltering}
Li, M.; Zhang, Y.; He, S.; Li, Z.; Zhao, H.; Wang, J.; Cheng, N.; and Zhou, T.
  2024{\natexlab{a}}.
\newblock {Superfiltering: Weak-to-Strong Data Filtering for Fast
  Instruction-Tuning}.
\newblock In \emph{Proceedings of the 62nd Annual Meeting of the Association
  for Computational Linguistics (Volume 1: Long Papers)}, 14255--14273.
  Bangkok, Thailand: Association for Computational Linguistics.

\bibitem[{Li et~al.(2024{\natexlab{b}})Li, Zhang, Li, Chen, Chen, Cheng, Wang,
  Zhou, and Xiao}]{li2024fromquantity}
Li, M.; Zhang, Y.; Li, Z.; Chen, J.; Chen, L.; Cheng, N.; Wang, J.; Zhou, T.;
  and Xiao, J. 2024{\natexlab{b}}.
\newblock {From Quantity to Quality: Boosting LLM Performance with Self-Guided
  Data Selection for Instruction Tuning}.
\newblock In \emph{Proceedings of the 2024 Conference of the North American
  Chapter of the Association for Computational Linguistics: Human Language
  Technologies (Volume 1: Long Papers)}, 7602--7635.

\bibitem[{Li et~al.(2024{\natexlab{c}})Li, Hui, Xia, Yang, Yang, Zhang, Si,
  Chen, Liu, Liu, Huang, and Li}]{li2024nuggets}
Li, Y.; Hui, B.; Xia, X.; Yang, J.; Yang, M.; Zhang, L.; Si, S.; Chen, L.-H.;
  Liu, J.; Liu, T.; Huang, F.; and Li, Y. 2024{\natexlab{c}}.
\newblock {One-Shot Learning as Instruction Data Prospector for Large Language
  Models}.
\newblock In \emph{Proceedings of the 62nd Annual Meeting of the Association
  for Computational Linguistics (Volume 1: Long Papers)}.

\bibitem[{Lin, Hilton, and Evans(2022)}]{lin2022truthfulqa}
Lin, S.; Hilton, J.; and Evans, O. 2022.
\newblock {TruthfulQA: Measuring How Models Mimic Human Falsehoods}.
\newblock In \emph{Proceedings of the 60th Annual Meeting of the Association
  for Computational Linguistics (Volume 1: Long Papers)}, 3214--3252.

\bibitem[{Liu et~al.(2024)Liu, Zeng, He, Jiang, and He}]{liu2024makesgood}
Liu, W.; Zeng, W.; He, K.; Jiang, Y.; and He, J. 2024.
\newblock {What Makes Good Data for Alignment? A Comprehensive Study of
  Automatic Data Selection in Instruction Tuning}.
\newblock In \emph{ICLR}.

\bibitem[{Longpre et~al.(2024)Longpre, Mahari, Chen, Obeng-Marnu, Sileo,
  Brannon, Muennighoff, Khazam, Kabbara, Perisetla, Wu, Shippole, Bollacker,
  Wu, Villa, Pentland, and Hooker}]{longpre2024provenance}
Longpre, S.; Mahari, R.; Chen, A.; Obeng-Marnu, N.; Sileo, D.; Brannon, W.;
  Muennighoff, N.; Khazam, N.; Kabbara, J.; Perisetla, K.; Wu, X.; Shippole,
  E.; Bollacker, K.; Wu, T.; Villa, L.; Pentland, S.; and Hooker, S. 2024.
\newblock {A Large-Scale Audit of Dataset Licensing and Attribution in AI}.
\newblock \emph{Nature Machine Intelligence}, 6: 975--987.

\bibitem[{Lu et~al.(2024)Lu, Yuan, Yuan, Lin, Lin, Tan, Zhou, and
  Zhou}]{lu2024instag}
Lu, K.; Yuan, H.; Yuan, Z.; Lin, R.; Lin, J.; Tan, C.; Zhou, C.; and Zhou, J.
  2024.
\newblock {\#InsTag: Instruction Tagging for Analyzing Supervised Fine-tuning
  of Large Language Models}.
\newblock In \emph{ICLR}.

\bibitem[{Ouyang et~al.(2022)Ouyang, Wu, Jiang, Almeida, Wainwright, Mishkin,
  Zhang, Agarwal, Slama, Ray, Schulman, Hilton, Kelton, Miller, Simens, Askell,
  Welinder, Christiano, Leike, and Lowe}]{ouyang2022training}
Ouyang, L.; Wu, J.; Jiang, X.; Almeida, D.; Wainwright, C.; Mishkin, P.; Zhang,
  C.; Agarwal, S.; Slama, K.; Ray, A.; Schulman, J.; Hilton, J.; Kelton, F.;
  Miller, L.; Simens, M.; Askell, A.; Welinder, P.; Christiano, P.; Leike, J.;
  and Lowe, R. 2022.
\newblock {Training Language Models to Follow Instructions with Human
  Feedback}.
\newblock In \emph{NeurIPS}, 27730--27744.

\bibitem[{Pathmanathan et~al.(2026)Pathmanathan, Sehwag, Panaitescu-Liess,
  Chiang, and Huang}]{zhang2024advbdgen}
Pathmanathan, P.; Sehwag, U.~M.; Panaitescu-Liess, M.-A.; Chiang, C.-Y.~J.; and
  Huang, F. 2026.
\newblock {AdvBDGen: A Robust Framework for Generating Adaptive and Stealthy
  Backdoors in LLM Alignment}.
\newblock In \emph{Proceedings of the AAAI Conference on Artificial
  Intelligence}, volume~40, 37820--37829.

\bibitem[{Peng et~al.(2023)Peng, Li, He, Galley, and Gao}]{peng2023instruction}
Peng, B.; Li, C.; He, P.; Galley, M.; and Gao, J. 2023.
\newblock {Instruction Tuning with GPT-4}.
\newblock arXiv:2304.03277.

\bibitem[{Peng et~al.(2025)Peng, Cui, Huang, Yang, and Ray}]{peng2025cweval}
Peng, J.; Cui, L.; Huang, K.; Yang, J.; and Ray, B. 2025.
\newblock {CWEval: Outcome-driven Evaluation on Functionality and Security of
  LLM Code Generation}.
\newblock arXiv:2501.08200.

\bibitem[{Qi et~al.(2024)Qi, Zeng, Xie, Chen, Jia, Mittal, and
  Henderson}]{qi2024finetuning}
Qi, X.; Zeng, Y.; Xie, T.; Chen, P.-Y.; Jia, R.; Mittal, P.; and Henderson, P.
  2024.
\newblock {Fine-tuning Aligned Language Models Compromises Safety, Even When
  Users Do Not Intend To!}
\newblock In \emph{ICLR}.

\bibitem[{{Qwen Team}(2024)}]{yang2024qwen25}
{Qwen Team}. 2024.
\newblock {Qwen2.5 Technical Report}.
\newblock arXiv:2412.15115.

\bibitem[{{Qwen Team}(2025)}]{qwen2025qwen3}
{Qwen Team}. 2025.
\newblock {Qwen3 Technical Report}.
\newblock arXiv:2505.09388.

\bibitem[{R\"{o}ttger et~al.(2024)R\"{o}ttger, Kirk, Vidgen, Attanasio,
  Bianchi, and Hovy}]{rottger2024xstest}
R\"{o}ttger, P.; Kirk, H.~R.; Vidgen, B.; Attanasio, G.; Bianchi, F.; and Hovy,
  D. 2024.
\newblock {XSTest: A Test Suite for Identifying Exaggerated Safety Behaviours
  in Large Language Models}.
\newblock In \emph{NAACL-HLT}, 5377--5400.

\bibitem[{Shen et~al.(2025)Shen, Chen, Das, and Chen}]{shen2025sealsafety}
Shen, H.; Chen, P.-Y.; Das, P.; and Chen, T. 2025.
\newblock {SEAL: Safety-enhanced Aligned LLM Fine-tuning via Bilevel Data
  Selection}.
\newblock In \emph{ICLR}.

\bibitem[{Shu et~al.(2023)Shu, Wang, Zhu, Geiping, Xiao, and
  Goldstein}]{shu2023exploitabilityinstruction}
Shu, M.; Wang, J.; Zhu, C.; Geiping, J.; Xiao, C.; and Goldstein, T. 2023.
\newblock {On the Exploitability of Instruction Tuning}.
\newblock In \emph{NeurIPS}.

\bibitem[{Souly et~al.(2025)Souly, Rando, Chapman, Davies, Hasircioglu,
  Shereen, Mougan, Mavroudis, Jones, Hicks, Carlini, Gal, and
  Kirk}]{souly2025nearconstant}
Souly, A.; Rando, J.; Chapman, E.; Davies, X.; Hasircioglu, B.; Shereen, E.;
  Mougan, C.; Mavroudis, V.; Jones, E.; Hicks, C.; Carlini, N.; Gal, Y.; and
  Kirk, R. 2025.
\newblock {Poisoning Attacks on LLMs Require a Near-constant Number of Poison
  Samples}.
\newblock arXiv:2510.07192.

\bibitem[{Taori et~al.(2023)Taori, Gulrajani, Zhang, Dubois, Li, Guestrin,
  Liang, and Hashimoto}]{taori2023stanford}
Taori, R.; Gulrajani, I.; Zhang, T.; Dubois, Y.; Li, X.; Guestrin, C.; Liang,
  P.; and Hashimoto, T.~B. 2023.
\newblock {Stanford Alpaca: An Instruction-Following LLaMA Model}.
\newblock GitHub repository, accessed 2026-06-04.

\bibitem[{Wallace et~al.(2021)Wallace, Zhao, Feng, and
  Singh}]{wallace2021concealed}
Wallace, E.; Zhao, T.; Feng, S.; and Singh, S. 2021.
\newblock {Concealed Data Poisoning Attacks on NLP Models}.
\newblock In \emph{NAACL-HLT}, 139--150.

\bibitem[{Wan et~al.(2023)Wan, Wallace, Shen, and
  Klein}]{wan2023poisoninglanguage}
Wan, A.; Wallace, E.; Shen, S.; and Klein, D. 2023.
\newblock {Poisoning Language Models During Instruction Tuning}.
\newblock In \emph{ICML}, 35413--35425.

\bibitem[{Wang et~al.(2023)Wang, Kordi, Mishra, Liu, Smith, Khashabi, and
  Hajishirzi}]{wang2023selfinstruct}
Wang, Y.; Kordi, Y.; Mishra, S.; Liu, A.; Smith, N.~A.; Khashabi, D.; and
  Hajishirzi, H. 2023.
\newblock {Self-Instruct: Aligning Language Models with Self-Generated
  Instructions}.
\newblock In \emph{ACL}, 13484--13508.

\bibitem[{Xia et~al.(2024)Xia, Malladi, Gururangan, Arora, and
  Chen}]{xia2024lessselecting}
Xia, M.; Malladi, S.; Gururangan, S.; Arora, S.; and Chen, D. 2024.
\newblock {LESS: Selecting Influential Data for Targeted Instruction Tuning}.
\newblock In \emph{ICML}, 54104--54132.

\bibitem[{Xu et~al.(2024{\natexlab{a}})Xu, Sun, Zheng, Geng, Zhao, Feng, Tao,
  Lin, and Jiang}]{xu2023wizardlm}
Xu, C.; Sun, Q.; Zheng, K.; Geng, X.; Zhao, P.; Feng, J.; Tao, C.; Lin, Q.; and
  Jiang, D. 2024{\natexlab{a}}.
\newblock {WizardLM}: Empowering Large Pre-Trained Language Models to Follow
  Complex Instructions.
\newblock In \emph{The Twelfth International Conference on Learning
  Representations}.

\bibitem[{Xu et~al.(2024{\natexlab{b}})Xu, Ma, Wang, Xiao, and
  Chen}]{xu2024instructionsas}
Xu, J.; Ma, M.; Wang, F.; Xiao, C.; and Chen, M. 2024{\natexlab{b}}.
\newblock {Instructions as Backdoors: Backdoor Vulnerabilities of Instruction
  Tuning for Large Language Models}.
\newblock In \emph{Proceedings of the 2024 Conference of the North American
  Chapter of the Association for Computational Linguistics: Human Language
  Technologies (Volume 1: Long Papers)}, 3111--3126. Association for
  Computational Linguistics.

\bibitem[{Yan et~al.(2024)Yan, Yadav, Li, Chen, Tang, Wang, Srinivasan, Ren,
  and Jin}]{yan2024backdooringinstruction}
Yan, J.; Yadav, V.; Li, S.; Chen, L.; Tang, Z.; Wang, H.; Srinivasan, V.; Ren,
  X.; and Jin, H. 2024.
\newblock {Backdooring Instruction-Tuned Large Language Models with Virtual
  Prompt Injection}.
\newblock In \emph{NAACL-HLT}, 6065--6086.

\bibitem[{Zhang et~al.(2025)Zhang, Wang, Du, Zhang, Tu, and
  Chu}]{zhang2025surveydata}
Zhang, B.; Wang, J.; Du, Q.; Zhang, J.; Tu, Z.; and Chu, D. 2025.
\newblock {A Survey on Data Selection for LLM Instruction Tuning}.
\newblock \emph{Journal of Artificial Intelligence Research}, 83.

\bibitem[{Zheng et~al.(2023)Zheng, Chiang, Sheng, Zhuang, Wu, Zhuang, Lin, Li,
  Li, Xing, Zhang, Gonzalez, and Stoica}]{zheng2023judgingllm}
Zheng, L.; Chiang, W.-L.; Sheng, Y.; Zhuang, S.; Wu, Z.; Zhuang, Y.; Lin, Z.;
  Li, Z.; Li, D.; Xing, E.~P.; Zhang, H.; Gonzalez, J.~E.; and Stoica, I. 2023.
\newblock {Judging LLM-as-a-Judge with MT-Bench and Chatbot Arena}.
\newblock In \emph{Advances in Neural Information Processing Systems 36
  (NeurIPS 2023), Datasets and Benchmarks Track}.

\bibitem[{Zhou et~al.(2023)Zhou, Liu, Xu, Iyer, Sun, Mao, Ma, Efrat, Yu, Yu,
  Zhang, Ghosh, Lewis, Zettlemoyer, and Levy}]{zhou2023lima}
Zhou, C.; Liu, P.; Xu, P.; Iyer, S.; Sun, J.; Mao, Y.; Ma, X.; Efrat, A.; Yu,
  P.; Yu, L.; Zhang, S.; Ghosh, G.; Lewis, M.; Zettlemoyer, L.; and Levy, O.
  2023.
\newblock {LIMA: Less Is More for Alignment}.
\newblock In \emph{NeurIPS}, 55006--55021.

\bibitem[{Zhou et~al.(2024)Zhou, Qiang, Zade, Roshani, Khanduri, Zytko, and
  Zhu}]{qiang2024learningpoison}
Zhou, X.; Qiang, Y.; Zade, S.~Z.; Roshani, M.~A.; Khanduri, P.; Zytko, D.; and
  Zhu, D. 2024.
\newblock {Learning to Poison Large Language Models for Downstream
  Manipulation}.
\newblock arXiv:2402.13459.

\end{thebibliography}
}

\clearpage
\appendix
\setcounter{figure}{0}
\renewcommand{\thefigure}{S\arabic{figure}}
\setcounter{table}{0}
\renewcommand{\thetable}{S\arabic{table}}
\setcounter{algorithm}{0}
\renewcommand{\thealgorithm}{S\arabic{algorithm}}
\section*{Technical Supplement}
This supplement supplies the technical and reproducibility details behind the main paper: an artifact map, frozen configurations, run counts, evaluation and statistical protocols, the strict random-selection control, CONCERT pseudocode, selector--payload measurements, budget and poison-rate sweeps, and complete proofs. The main paper remains self-contained. Proposition numbering follows the main paper, except that the monotonicity result appears here as Proposition~S1 because the main text reports only its empirical trends.

\section*{Reproducibility Summary}

\paragraph{Artifact identity and integrity.}
The companion archive \path{code_and_data_supplement.zip}, provided with this preprint's ancillary files, is 1.65~MiB and contains 182 regular files. Its SHA-256 digest is
\begin{center}
\small
\texttt{2824e80e2fdb5914}\allowbreak\texttt{44f067d714b5816e}\allowbreak\texttt{dd79eb49422c93d8}\allowbreak\texttt{98c1a18a60f10cb3}.
\end{center}
The archive is preserved byte-for-byte from the frozen reproducibility release; its original documentation and manifest remain unchanged. It contains preprocessing, selection, training, evaluation, and statistical code; frozen configurations and prompts; generated candidate pools; per-seed and aggregate tables; data-lineage manifests; and \path{MANIFEST_SHA256.txt}, which authenticates every packaged file. The build report records successful checks of the listed headline numbers, 51 Python files, 14 JSON files, 69 CSV files, 15 JSONL files, two focused tests, the three-CWE candidate audit, anonymity, author-machine paths, private-key patterns, and high-confidence secret patterns. No \texttt{import openai} statement is present, and the original artifact code is accompanied by an MIT \path{LICENSE}.

\paragraph{What is and is not redistributed.}
Generated candidates, score/selection tables, small derived results, the sanitized 500-record Qwen AMRS over-refusal pool, exact prompts, and output hashes are included. Third-party model weights and benchmark corpora, LoRA adapters, full benchmark completions, and some non-Qwen generated pools are not redistributed because of the upload limit and upstream licenses. Public identifiers, checkpoint fingerprints, schemas, configurations, and lineage hashes are supplied instead. The hosted over-refusal judge identifier was not preserved and is not guessed. The deterministic code detector and TruthfulQA MC1 therefore provide the reproducible primary metrics; the hosted semantic audits are corroborative.

\paragraph{Claim-to-file map.}
The principal audit entry points are:
\begin{itemize}
  \item false facts: \path{data/derived/false_facts/}, with the 32B paired results in \path{EA_multiseed_32b.csv} and \path{EA_multiseed_32b_ci.csv};
  \item insecure code: \path{data/derived/code_vulnerability/}, including the three-seed attack tables;
  \item CONCERT: \path{data/derived/concert/}, including fixed-budget selection, partner ablation, supplier-audit, and three-seed defense tables;
  \item adaptive reranking: \path{data/derived/adaptive/} and \path{data/adaptive_candidates/};
  \item over-refusal: \path{data/derived/over_refusal/}, \path{data/released/over_refusal/}, and \path{prompts/over_refusal/}.
\end{itemize}
The standard-library checker \path{code/statistics/reproduce_reported_numbers.py} recomputes the included paper-facing means, sample standard deviations, the 32B confidence interval, and over-refusal summaries without inference or training. The complete map is in \path{docs/REPRODUCIBILITY.md}.

\section*{Frozen Experimental Configuration}

\paragraph{Models and checkpoint identity.}
The public checkpoints are Qwen2.5-7B-Instruct, Meta-Llama-3-8B-Instruct, Qwen3-4B/8B/14B/32B, and OFA-Sys/InsTagger. Exact Qwen3 repository commits were retained: \texttt{2c54d5a09e7e} (4B), \texttt{26028140be3e} (8B), \texttt{6b837d17776d} (14B), and \texttt{d64d397e6690} (32B). Repository commits were not preserved for Qwen2.5-7B, Llama-3-8B, or InsTagger; for them, the SHA-256 fingerprints of \path{config.json} and the safetensors index are authoritative. Full commits and fingerprints are in \path{configs/model_checkpoints.csv}.

\begin{table*}[t]
\centering
\small
\begin{tabular}{@{}p{0.15\textwidth}p{0.15\textwidth}p{0.11\textwidth}p{0.23\textwidth}p{0.19\textwidth}@{}}
\toprule
Experiment & Base model & LoRA targets & Optimization & Seeds and compute \\
\midrule
7B false facts & Qwen2.5-7B-Instruct & q/v & 3 epochs; length 1024; LR $2\times10^{-4}$; batch 1, accumulation 8; linear & 7/13/42; one GPU; effective batch 8 \\
32B false facts & Qwen3-32B & q/k/v/o & 3 epochs; length 1024; LR $2\times10^{-4}$; batch 1, accumulation 8; linear; FSDP full shard & 7/13/42; two H100 80-GB GPUs; effective batch 16 \\
CodeVuln and CONCERT & Qwen3-32B & q/v & 3 epochs; length 1536; LR $2\times10^{-4}$; batch 1, accumulation 8; cosine; ZeRO-3 CPU offload & 7/13/21; one H100 80-GB GPU; effective batch 8 \\
Over-refusal & Qwen2.5-7B and Llama-3-8B & q/v & 3 epochs; length 2048; LR $2\times10^{-4}$; batch 2, accumulation 4; cosine & Qwen 7/13/42; Llama 7; effective batch 8 \\
\bottomrule
\end{tabular}
\caption{Final fine-tuning configurations. All rows use LoRA rank 8, $\alpha=16$, dropout 0.05, AdamW, warmup ratio 0.03, zero weight decay, and bf16. H100 memory follows the supplied cluster inventory; unretained job allocations are identified in text.}
\label{tab:supp-train-config}
\end{table*}

The final settings are frozen in \path{configs/training_configs.json}; the exact DeepSpeed file is \path{code/training/ds_zero3_cpu_offload_qwen32.json}. Development did not retain a complete exhaustive hyperparameter-search ledger, so we do not claim one. The explicitly varied settings are the adaptive candidate budget $K\in\{4,8,16,32\}$, over-refusal AMRS $K=10$ with temperatures $\{0.5,0.7,0.9,1.1\}$, adaptive-code temperatures $\{0.35,0.55,0.75,0.95\}$, selection budgets, poison rates, models, and selectors reported in the paper and artifact.

\paragraph{Hardware and software.}
The supplied H100-cluster inventory specifies two 32-core 2.8-GHz Intel Xeon Gold 8462Y+ CPUs, 2,048~GB node memory, and eight NVIDIA H100 80-GB GPUs connected by NVLink per node. Retained job-allocation records identify the H100 GPU family and show that the 32B false-fact jobs allocated two GPUs and the code/CONCERT jobs one, but do not retain a node identifier; we therefore report the CPU, memory, and eight-GPU topology as cluster inventory rather than claiming the exact allocated node. The adaptive selection-only jobs used one A800. The historical Llama over-refusal run explicitly records one A100 80-GB GPU; the Qwen over-refusal GPU allocation was not retained.

The recorded stack is Python~3.11.5, PyTorch~2.1.2 with CUDA~11.8, torchvision~0.16.2, torchaudio~2.1.2, Triton~2.1.0, Transformers~4.46.3 (base) and 4.51.0 (selector/adaptive override), PEFT~0.13.2, Accelerate~0.34.2, Datasets~2.21.0, DeepSpeed~0.13.5, NumPy~1.26.2, pandas~2.2.0, tokenizers~0.20.3, and pyarrow~23.0.1. Frozen requirements are under \path{environment/}.

\section*{Run Inventory and Randomness}

\begin{table*}[t]
\centering
\small
\begin{tabular}{@{}p{0.25\textwidth}p{0.12\textwidth}p{0.14\textwidth}p{0.36\textwidth}@{}}
\toprule
Result family & Training runs & Seeds & Repetition and uncertainty \\
\midrule
7B false-fact selector comparison & 54 & 7/13/42 & Six selectors $\times$ three conditions $\times$ three paired seeds \\
32B false-fact downstream & 12 & 7/13/42 & Four conditions $\times$ three paired seeds; Student-$t$ interval and Holm correction \\
Strict equal-size random control & 1 & 7 & One additional random-$m200$ SFT; clean and IFD seed-7 runs are reused \\
CodeVuln attack & 6 & 7/13/21 & Three attack plus three paired clean runs; mean $\pm$ sample SD \\
CONCERT facts & 3 & 7 & Clean/attack/defense single-seed condition set; descriptive \\
CONCERT code defense & 3 & 7/13/21 & Three defense runs; mean $\pm$ sample SD \\
Over-refusal Qwen & 15 & 7/13/42 & Five conditions $\times$ three seeds; descriptive mean \\
Over-refusal Llama & 4 & 7 & Four conditions; single-seed confirmation \\
Scale, transfer, and low rate & 0 & 7 & Frozen selection analyses; no SFT and no inferential interval \\
Adaptive reranking & 0 & 7 & One generator run; the $K$ sweep is nested post-processing, not four repetitions \\
\bottomrule
\end{tabular}
\caption{Run inventory. Counts are training runs; deterministic frozen-pool analyses are listed as zero SFT runs.}
\label{tab:supp-runs}
\end{table*}

Random seeds are passed to Python, NumPy, PyTorch, candidate generation, and the training launchers. Top-$k$ ties use average percentile ranks followed by a stable record-key tie-break whenever exact cardinality is required. Per-condition seeds and run counts are machine-readable in \path{configs/experiment_runs.csv}.

\section*{Evaluation and Statistical Protocols}

\paragraph{Metrics and judges.}
TruthfulQA MC1 is the primary false-fact metric and uses no LLM judge. MMLU measures general language capability. HumanEval measures functional code correctness. The primary insecure-code evaluator is the deterministic \texttt{targeted-cwe-v1.0.2} AST/regex detector: it applies target-specific binary rules, has no temperature and no probability threshold, and uses CWE-078, CWE-295, and CWE-502 in the main downstream aggregate. The local semantic audit is corroborative because its exact interactive serving snapshot was not programmatically recorded.

The over-refusal prompt is packaged at \path{prompts/over_refusal/gpt_refusal_judge_prompt.txt}. Decoding temperature is 0; \texttt{FULL\_REFUSAL} and \texttt{PARTIAL\_REFUSAL} are positive, and there is no probability threshold. The concrete hosted judge model identifier was not preserved and is not guessed. HumanAudit-400 retained binary human--human $\kappa=0.988$ and judge-versus-human precision/recall/F1 of 0.966/0.983/0.974. Only aggregate confusion and sample tables were retained; raw double annotations and annotator recruitment, compensation, and IRB metadata are not packaged, so we do not claim those details can be independently reconstructed.

\paragraph{Uncertainty and multiplicity.}
The CWE-078/295/502 code attack $+56.67\pm9.62$ pp and HumanEval $-0.41\pm1.27$ pp are mean $\pm$ sample SD across seeds 7/13/21. The CONCERT HumanEval change $+0.20\pm0.35$ pp is the same statistic over the three defense-minus-attack deltas. For 32B false facts, $[-0.079,-0.045]$ is a two-sided Student-$t$ 95\% confidence interval for the mean paired held-out MC1 change with $n=3$ and $df=2$. The raw two-sided $p=0.00409$ becomes $p=0.04499$ after Holm correction over 12 non-clean headline truthfulness tests.

\paragraph{Strict equal-size random control.}
The strict seed-7 control holds the training size and poison count fixed at 825 and 200. The IFD condition contains 200 selector-chosen poison records; the random condition draws 200 poison identities from the same frozen 500-record poison pool and uses the same 625-record clean background. It therefore matches counts and clean context, not the exact poison identities. It is distinct from the three-seed unconcentrated condition, which contains all 500 poison records in the 5,500-record incoming pool and has mean MC1 change $-0.03448$ with Holm-adjusted $p=0.28494$.

\begin{table*}[t]
\centering
\small
\begin{tabular}{@{}lcccccc@{}}
\toprule
Condition & MC1 & $\Delta$MC1 & Gen. True & $\Delta$True & True$\times$Info & $\Delta$(T$\times$I) \\
\midrule
Clean & 0.3621 & 0 & 0.6793 & 0 & 0.6552 & 0 \\
IFD-selected $m=200$ & 0.3069 & $-0.0552$ & 0.6414 & $-0.0379$ & 0.6172 & $-0.0379$ \\
Random $m=200$ & 0.3069 & $-0.0552$ & 0.6310 & $-0.0483$ & 0.6069 & $-0.0483$ \\
\bottomrule
\end{tabular}
\caption{Strict equal-size random control on held-out TruthfulQA. IFD selection is not more harmful on any of the three reported truthfulness summaries, and held-out MC1 is identical. This control therefore does not identify an additional selector-specific causal effect. Source: \path{data/derived/false_facts/ED_random_vs_selector_32b.csv}.}
\label{tab:supp-strict-random}
\end{table*}

\paragraph{Supplier audit and partner choice.}
Both Holm FWER control and Benjamini--Hochberg FDR control use level 0.05; the spread study uses 200 trials per cell and 1,000 source-exchangeability null trials. At six suppliers, BH detects 48.25\%/33.83\% and Holm 28.5\%/17.25\% for facts/code. The aggregate audit summary is packaged, but the standalone null-draw file and its original \path{b1_config.json} were not retained. Partner selection computes Pearson correlation on tie-averaged percentile ranks, which is Spearman rank correlation of the underlying scores. For false facts the anchor DEITA is compared with IFD, AlpaGasus, and LESS, selecting AlpaGasus; for code it is compared with IFD, selecting IFD. No poison labels enter this choice.

\paragraph{Code scope and denominators.}
All reported insecure-code results use CWE-078/295/502. At the selector stage, $N=1{,}488$, $m=93$, and $k=223$: DEITA selects $93/223$ poison records ($E=6.67\times$), IFD $0/223$ ($E=0$), Borda $7/223$ ($E=0.50\times$), and CONCERT-minimax $0/223$ ($E=0$). The downstream attack counts are $56/90$, $56/90$, and $41/90$ across seeds 7/13/21, with $0/90$ for clean and CONCERT on every seed. These downstream values audit the existing checkpoints on the stated three-CWE evaluator rather than retraining those checkpoints on the reduced pool. In the adaptive code test, consensus selects no CWE-078/295/502 poison at every tried $K$.

\section*{CONCERT Pseudocode}

\begin{algorithm}[H]
\caption{CONCERT rank-intersection selection}
\label{alg:concert}
\begin{algorithmic}[1]
\REQUIRE pool $\mathcal{D}$, supplier ids $z$, anchor selector $f_a$, candidate selectors $\mathcal{F}$, budget $k$
\ENSURE supplier audit; fixed-budget training set $\mathcal{T}$; selection + input hashes
\STATE run top-$k$ under $f_a$; count selections $X_s$ per supplier
\STATE for each $s$: exact hypergeometric $p_s$; apply Holm / BH at 0.05 $\rightarrow$ alarms
\STATE convert $f_a$ and each $f_j\in\mathcal{F}$ to tie-averaged percentile ranks
\STATE $j^\star \leftarrow \arg\min_j \mathrm{corr}(r_a, r_j)$ \hfill(Pearson on ranks; no labels)
\STATE $g(i) \leftarrow \max\{r_a(i), r_{j^\star}(i)\}$ for all $i$
\STATE $\mathcal{T} \leftarrow$ the $k$ records with smallest $g$ (deterministic tie-break)
\RETURN audit alarms, $\mathcal{T}$, selection + input hashes
\end{algorithmic}
\end{algorithm}

Candidate-selector ranks may be precomputed. If they are not cached, choosing from $J$ candidate partners requires $J$ scoring passes; after $j^\star$ has been fixed, deployment adds one partner-scoring pass relative to the anchor-only pipeline. Pearson correlation of tie-averaged percentile ranks is the Spearman rank correlation of the underlying selector scores. The fact candidate set is \{IFD, AlpaGasus, LESS\}, from which AlpaGasus is selected; the code candidate set is \{IFD\}. The $0.97$ H100-hours reported in the main text are for the selected AlpaGasus partner, while the $1.04$ H100-hour total includes both that partner and the $0.069$ H100-hour DEITA anchor.

\section*{Selector--Payload Compatibility Matrix}

\begin{figure*}[t]
\centering
\includegraphics[width=\textwidth]{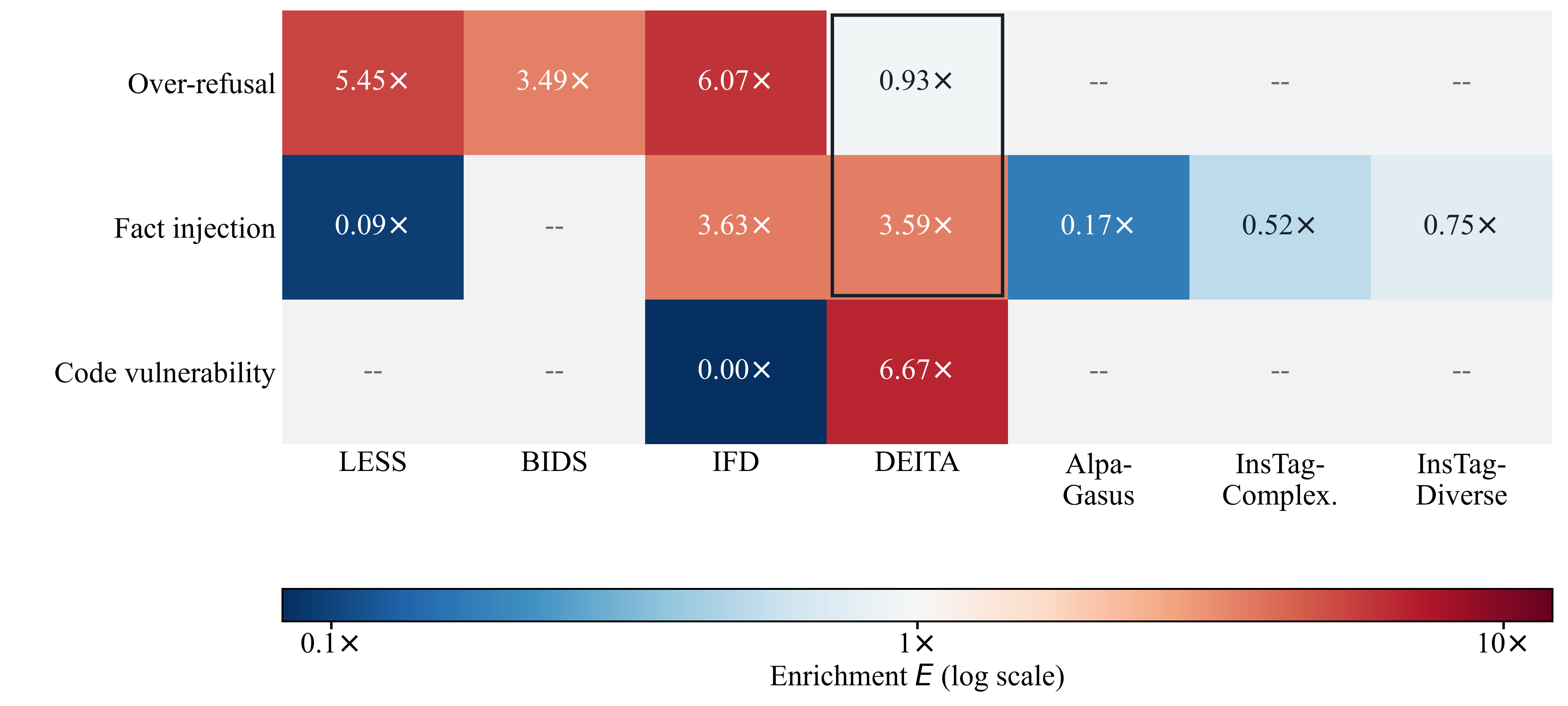}
\caption{Selector--payload compatibility. Cell values are selector-stage enrichment $E$ at Top-$15\%$, with a diverging scale centered at $1\times$; blank cells were not measured. Over-refusal is the Qwen2.5-7B seed-7 cross-selector slice (so LESS is $5.45\times$, versus the main three-seed mean $5.18\times$); false facts use the Qwen3-14B fixed pool; code is the seed-7 CWE-078/295/502 diagnostic above. Exact zero is displayed at the lower end of the color scale.}
\label{fig:supp-heatmap}
\end{figure*}

Figure~\ref{fig:supp-heatmap} gives the measured compatibility matrix referenced in the main text. Enrichment is measured at Top-$15\%$ under the finite-rate definition, and cells that were not measured are left blank rather than imputed. The measured cells show that immunity need not transfer across payloads: DEITA dilutes over-refusal poison but amplifies false facts and insecure code, whereas IFD amplifies over-refusal and false facts but dilutes the insecure-code payload. We make no claim about blank cells.

\begin{table}[t]
\centering
\small
\begin{tabular}{@{}llcc@{}}
\toprule
Payload & Selector & $q_f(15\%)$ & $E$ \\
\midrule
False facts & DEITA & 0.596 & $3.59\times$ \\
False facts & IFD & 0.624 & $3.63\times$ \\
False facts & AlpaGasus & 0 & $0.17\times$ \\
False facts & LESS & 0.008 & $0.09\times$ \\
Insecure code & DEITA & 1.000 & $6.67\times$ \\
Insecure code & IFD & 0 & $0\times$ \\
\bottomrule
\end{tabular}
\caption{The six labeled diagnostic cells used for the CREST separation claim. Here $q_f(b)$ is the fraction of labeled poison above the cutoff induced by selecting the Top-$b$ records from the clean-only pool; $S_p$ instead uses the realized mixed-pool Top-$k$ cutoff. Thus $q_f=0$ can coexist with nonzero finite-pool enrichment. This is a labeled diagnostic, not an unlabeled online detector. Code values use CWE-078/295/502.}
\label{tab:supp-qf}
\end{table}

The six $q_f$ values are frozen here for audit; the final code/data archive does not contain their legacy source slice as a standalone derived CSV.

\paragraph{Insecure-code scope.}
The insecure-code evaluation uses CWE-078 (command injection), CWE-295 (TLS validation), and CWE-502 (unsafe deserialization). Starting from the frozen seed-7 selector rankings, the pool contains $1{,}395$ clean records and $93$ poison records, giving $N=1{,}488$, $m=93$, and Top-$15\%$ budget $k=223$. DEITA selects all $93$ poison records ($93/223$, $E=6.67\times$), IFD selects none ($0/223$, $E=0$), Borda selects $7/223$ ($E=0.50\times$), and CONCERT-minimax selects none ($0/223$, $E=0$). Across seeds 7/13/21, the downstream attack--clean gap is $+56.67\pm9.62$ pp. The downstream checkpoints predate this scope correction, so these downstream values are a three-CWE audit of the existing checkpoints rather than results of retraining on a reduced pool.

\section*{Budget and Poison-Rate Sweeps}

\begin{figure}[t]
\centering
\includegraphics[width=\columnwidth]{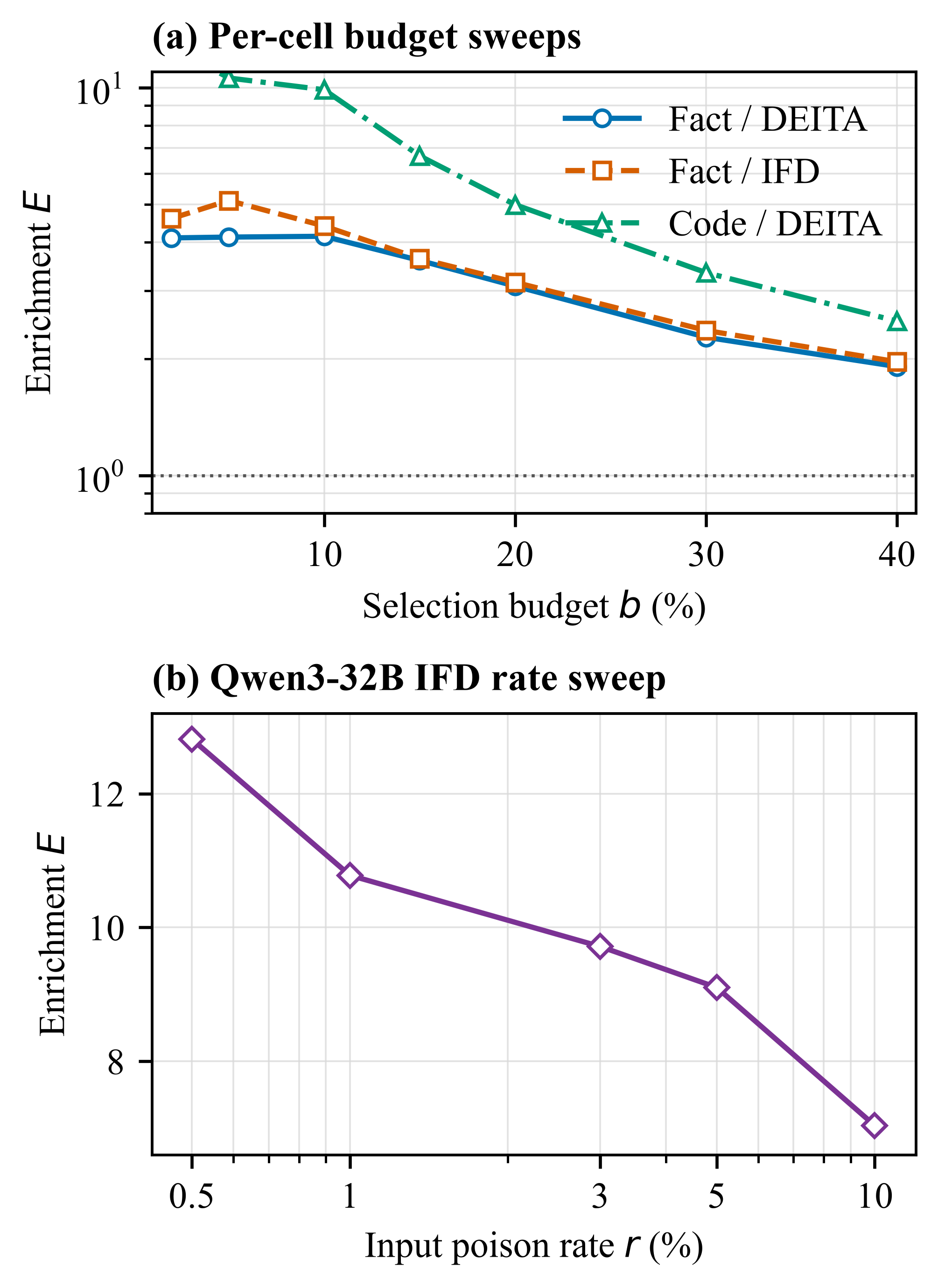}
\caption{Budget and poison-rate sweeps. (a) Observed finite-rate enrichment for the three measured vulnerable selector--payload cells; the code curve uses CWE-078/295/502, as does the downstream audit. (b) Qwen3-32B IFD enrichment for false-fact poison as the input poison rate varies. Line styles and markers make the curves distinguishable without color.}
\label{fig:supp-sweeps}
\end{figure}

Figure~\ref{fig:supp-sweeps} supplies the per-cell curves referenced in the main paper. Over the prespecified practical range $b=10$--$40\%$, enrichment decreases monotonically with budget for the three vulnerable cells shown. The $2$--$10\%$ points expose the mild finite-sample reversals discussed in the main text rather than hiding them. In the rate sweep, reducing the input poison fraction from $10\%$ to $0.5\%$ increases enrichment from $7.04\times$ to $12.81\times$.

\section*{Supplementary Proofs}

Throughout, the candidate pool has $N$ records, of which $m$ are poison, giving poisoning rate $r=m/N$. A selector score together with a deterministic stable record key induces a total order, and top-$k$ selection keeps the first $k=bN$ records. We write $\{s\ge\tau_b\}$ as shorthand for membership in this unique Top-$k$ set; this definition also covers discrete selector scores and ties. For the realized finite pool, write
\[
S_c=\frac{|\mathcal{T}\cap\mathcal{D}_{\mathrm{clean}}|}{N-m},\qquad
S_p=\frac{|\mathcal{T}\cap\mathcal{D}_{\mathrm{poison}}|}{m}
\]
for the empirical clean and poison tail-survival fractions (equivalently, probabilities under a uniform draw from the corresponding finite subgroup), and
\[
E=\frac{|\mathcal{T}\cap\mathcal{D}_{\mathrm{poison}}|/k}{m/N}
\]
for the enrichment.

\paragraph{Proposition 1 (Upper-tail enrichment).}
\emph{Under fixed budget fraction $b$ and rate $r$, $E=S_p/\!\left(rS_p+(1-r)S_c\right)$. If $S_c>0$, the survival probabilities are locally stable, and $r|S_p/S_c-1|\ll1$, then $E\approx S_p/S_c$.}

\begin{proof}
Exactly $mS_p$ poison and $(N-m)S_c$ clean records reach the tail, so the selected set has size
\begin{align*}
k &= mS_p+(N-m)S_c,\\
b &= \frac{k}{N}=rS_p+(1-r)S_c .
\end{align*}
The selected poison count is $mS_p$, so
\[
E=\frac{mS_p/k}{m/N}=\frac{N S_p}{k}=\frac{S_p}{b}=\frac{S_p}{rS_p+(1-r)S_c}.
\]
The finite-rate result is an exact identity for the realized empirical fractions (and analogously for a population selection rule). Writing $A=S_p/S_c$ gives $E=A/[1+r(A-1)]$; hence the stated condition makes $E\approx A$. We do not assert an unconditional $r\to0$ limit while allowing the cutoff and survival probabilities to vary arbitrarily with $r$.
\end{proof}

\paragraph{Proposition S1 (Monotonicity).}
\emph{This statement is made in the supplement only; the main text reports the corresponding empirical trends. Fix clean and poison score distributions with $S_p>S_c$. Then (i) $E\to1$ as $b\to1$; (ii) at a fixed cutoff, $E$ is strictly decreasing in $r$; and (iii) if the poison density dominates the clean density across the upper tail in the monotone-likelihood-ratio (MLR) sense, then $E$ is non-increasing in $b$.}

\begin{proof}
(i) As $b\to1$, $\tau_b$ falls below the support of both distributions, so $S_p,S_c\to1$ and $E=S_p/(rS_p+(1-r)S_c)\to 1/(r+(1-r))=1$.

(ii) At a fixed cutoff, $S_p$ and $S_c$ are constants and
\[
E(r)=\frac{S_p}{S_c+r\,(S_p-S_c)} .
\]
Since $S_p>S_c$, the denominator is strictly increasing in $r$, so $E$ is strictly decreasing in $r$.

(iii) Let $f_c,f_p$ denote the clean and poison score densities and view $S_c,S_p$ as functions of the cutoff $\tau=\tau_b$. Differentiating the budget identity $b=rS_p+(1-r)S_c$ in $b$ gives
\[
1=\big(r f_p(\tau)+(1-r)f_c(\tau)\big)\left(-\frac{d\tau}{db}\right),
\]
and $E=S_p/b$ has derivative $\frac{dE}{db}=\frac{b\,S_p'(b)-S_p}{b^2}$ with $S_p'(b)=f_p(\tau)\left(-\tfrac{d\tau}{db}\right)=\dfrac{f_p(\tau)}{r f_p(\tau)+(1-r)f_c(\tau)}$. Hence $\frac{dE}{db}\le0$ iff $b\,f_p(\tau)\le S_p\big(rf_p(\tau)+(1-r)f_c(\tau)\big)$, i.e.
\[
\frac{f_p(\tau)}{rf_p(\tau)+(1-r)f_c(\tau)}\le\frac{S_p}{b}.
\]
The left side is the poison share of the density crossing $\tau$, and the right side is the poison share of the mass already above $\tau$. Under MLR the ratio $f_p/f_c$ is non-decreasing in the score, so the marginal poison share at the cutoff does not exceed the average poison share above it, which is exactly the inequality above. Therefore $E$ is non-increasing in $b$. Away from the MLR regime, in particular at very small budgets in finite samples, the inequality can fail locally, which is the source of the mild reversals reported in the main text.
\end{proof}

\paragraph{Proposition 2 (Fixed budget and scale invariance).}
\emph{With a deterministic tie-break, CONCERT-Consensus returns exactly $k$ records on any input. It depends only on within-selector percentile ranks; replacing any selector $f_j$ by $h\circ f_j$ for strictly increasing $h$ leaves the selection unchanged.}

\begin{proof}
Each record $i$ receives the minimax score $g(i)=\max\{r_a(i),r_{j^\star}(i)\}$. The deterministic tie-break induces a total order on $\{g(i)\}$, and the rule keeps the $k$ smallest elements of a totally ordered finite set, which is a set of size exactly $k$ for any $k\le N$.

A strictly increasing $h$ preserves the order of the scores of $f_j$, hence preserves the tie-averaged percentile ranks $r_j(\cdot)$. Since $g$ and the partner choice $j^\star=\arg\min_j\mathrm{corr}(r_a,r_j)$ depend on the scores only through these ranks, both are unchanged, and therefore so is the selected set $\mathcal{T}$. In particular selectors with incomparable score scales can be combined without calibration.
\end{proof}

\paragraph{Proposition 3 (Minimax intersection certificate).}
\emph{Let $t^\star$ be the largest minimax score among selected records. Every selected record satisfies $r_a(i)\le t^\star$ and $r_{j^\star}(i)\le t^\star$, so $\mathcal{T}$ lies inside the intersection of the two selectors' Top-$t^\star$ prefixes, and for any poison set $P$,
$|\mathcal{T}\cap P|\le\big|P\cap\{i:r_a(i)\le t^\star,\ r_{j^\star}(i)\le t^\star\}\big|$.}

\begin{proof}
For any selected $i$, $\max\{r_a(i),r_{j^\star}(i)\}=g(i)\le t^\star$ by the definition of $t^\star$, hence $r_a(i)\le t^\star$ and $r_{j^\star}(i)\le t^\star$. Thus $\mathcal{T}\subseteq\{i:r_a(i)\le t^\star\}\cap\{i:r_{j^\star}(i)\le t^\star\}$. Intersecting both sides with an arbitrary $P$ and taking cardinalities gives the stated bound. The statement is a set-containment guarantee. It does not by itself bound the number of poison records, the enrichment, or the downstream harm, and it does not certify poison-free selection. Any record whose two numerical percentile ranks are both at most $t^\star$ can still enter.
\end{proof}

\end{document}